\documentclass[11pt,twoside]{article}

\usepackage{asp2004}
\usepackage{epsf}
\usepackage{psfig}
\usepackage{lscape}

\begin{document}

\title{Understanding Infrared Galaxy Populations: the SWIRE Legacy Survey}   

\author{Michael Rowan-Robinson$^1$, Carol Lonsdale, Gene Smith, Jason Surace, Dave Shupe,
Maria Polletta, Brian Siana, Tom Babbedge, Seb Oliver, Ismael Perez-Fournon, Alberto Franceschini,
Alejandro Afonso Luis, David Clements,Payam Davoodi, Donovan Domingue, Andreas Efstathiou, Fan Fang, Duncan Farrah,
Dave Frayer, Evanthia Hatziminaoglou, Eduardo Gonzalez-Solares, Kevin Xu, Deborah Padgett, Mattia Vaccari  }
\affil{
$^1$ Astrophysics Group, Blackett Laboratory, Imperial College London,
Prince Consort Road, London SW7 2BZ, UK,\\
}

\begin{abstract} 
We discuss spectral energy distributions, photometric redshifts, redshift distributions,
luminosity functions, source-counts and the far infrared to optical luminosity ratio 
for sources in the SWIRE Legacy Survey.

The spectral energy distributions of selected SWIRE sources
are modelled in terms of a simple set of galaxy and quasar templates in the optical 
and near infrared, and with a set of dust emission templates (cirrus, M82 starburst,
Arp 220 starburst, and AGN dust torus) in the mid infrared.

The optical data, together with the IRAC 3.6 and 4.5 $\mu$m data, have been used
to determine photometric redshifts.  For galaxies with known spectroscopic redshifts
there is a notable improvement in the photometric redshift when the IRAC data are used,
with a reduction in the rms scatter from 10 $\%$ in (1+z) to 5 $\%$.  While further spectroscopic
data are needed to confirm this result, the prospect of determining good photometric 
redshifts for the 2 million extragalactic 
objects in SWIRE is excellent.  
  The distribution of the different
infrared sed types in the $L_{ir}/L_{opt}$ versus $L_{ir}$ plane, where $L_{ir}$ and $L_{opt}$
are the infrared and optical bolometric luminosities, is discussed.

Source-counts at 24, 70 and 160 $\mu$m are discussed, and luminosity functions at 3.6 and 24 $\mu$m are
presented.

\end{abstract}

\section{Introduction}

Infrared wavelengths hold the key to understanding the evolution of galaxies.  From the spectrum
of the extragalactic background radiation ( Hauser and Dwek 2001) we see that much of the light
emitted by stars is absorbed by dust and reemitted at mid and  far infrared wavelengths.  Only
by understanding infrared extragalactic populations can we hope to get a reliable census of the
star formation history of galaxies and estimate the fraction of dust-obscured AGN.

The SPITZER SWIRE Legacy Survey has detected over 2 million infrared galaxies in 49 sq deg
of sky (Lonsdale et al 2003, 2004).  This will be especially powerful for searches for rare objects
and for systematic studies of the link between star formation history and large-scale stucture.
Reliable catalogues have now been delivered to SSC for all the 6 survey areas (Surace et al 2005).

\section{Spectral energy distributions}

Rowan-Robinson et al (2005) reported work on optical associations, sed modelling, photometric redshifts and
redshift distributions for SWIRE sources.  
To model the seds of SWIRE sources we have used an approach similar to that of Rowan-Robinson et al (2004) for
ELAIS sources.  The optical data and near infrared data to 4.5 $\mu$m are fitted with one of
the 8 optical galaxy templates used in the photometric redshift code of Rowan-Robinson
(2003), 6 galaxy templates (E, Sab, Sbc, Scd, Sdm, sb) and 2 AGN templates (Rowan-Robinson et 
al 2004).  The mid and far infrared data are fitted by a mixture of the 4 infrared templates
used by Rowan-Robinson (2001) in models for infrared and submillimetre source-counts: cirrus, M82 starburst,
Arp 220 starburst and AGN dust torus.  For each of these templates we have detailed
radiative transfer models (cirrus: Eftstathiou and Rowan-Robinson 2003, M82 and Arp 220 starbursts: Eftstathiou et al 
2000, AGN dust tori: Rowan-Robinson 1995, Efstathiou and Rowan-Robinson 1995).  The approach here is to try to
understand the overall SPITZER galaxy population.

Here we report on subsequent work on some interesting 
samples.  Fig 1 (L) shows seds for a sample of SWIRE-ELAIS galaxies for which we have IRS spectroscopy
(Perez-Fournon et al, 2006, in prep., Hernan-Caballero et al 2006). Fig. 1 (R) shows seds for a sample of
SWIRE-SHADES sources detected at 850 $\mu$m with SCUBA (Clements et al, 2006, in prep.).
For the IRS sample, selected to be bright at 15 $\mu$m, we find that a large fraction of
the sources are dominated by an AGN dust torus in the mid-ir.  However an interesting subset
show PAH features or strong silicate absorption.
The 850 $\mu$m sources show a wide variety of ir templates, including luminous cirrus 
components, and both M82 and Arp 220 starbursts, as well as several cases dominated by
an AGN dust torus in the mid-ir.  We also find a wide range of redshifts, with a median redshift of 1.75
and 20$\%$ of the sample at z $<$ 1.

The striking features that emerge from this modelling are that the seds can at least broadly be understood in terms of 
a small number of infrared templates.  Radiative transfer codes have several parameters and even better fits could 
be found by using the full range of models of Efstathiou et al (2000), Eftstathiou and Rowan-Robinson (2003).
There is excellent agreement between the IRS and SWIRE data and good consistency with the template predictions.  

\begin{figure}
\plottwo{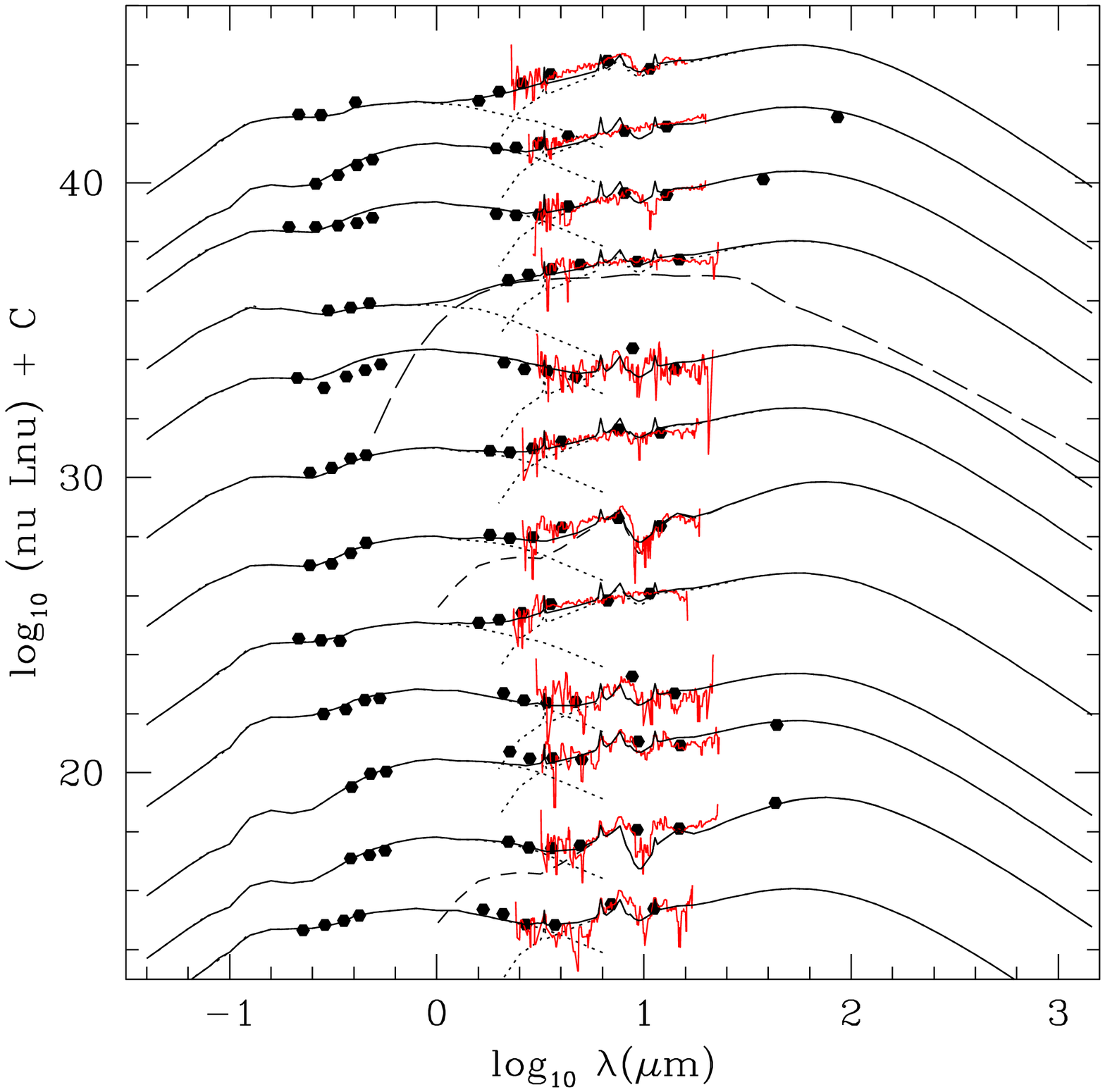}{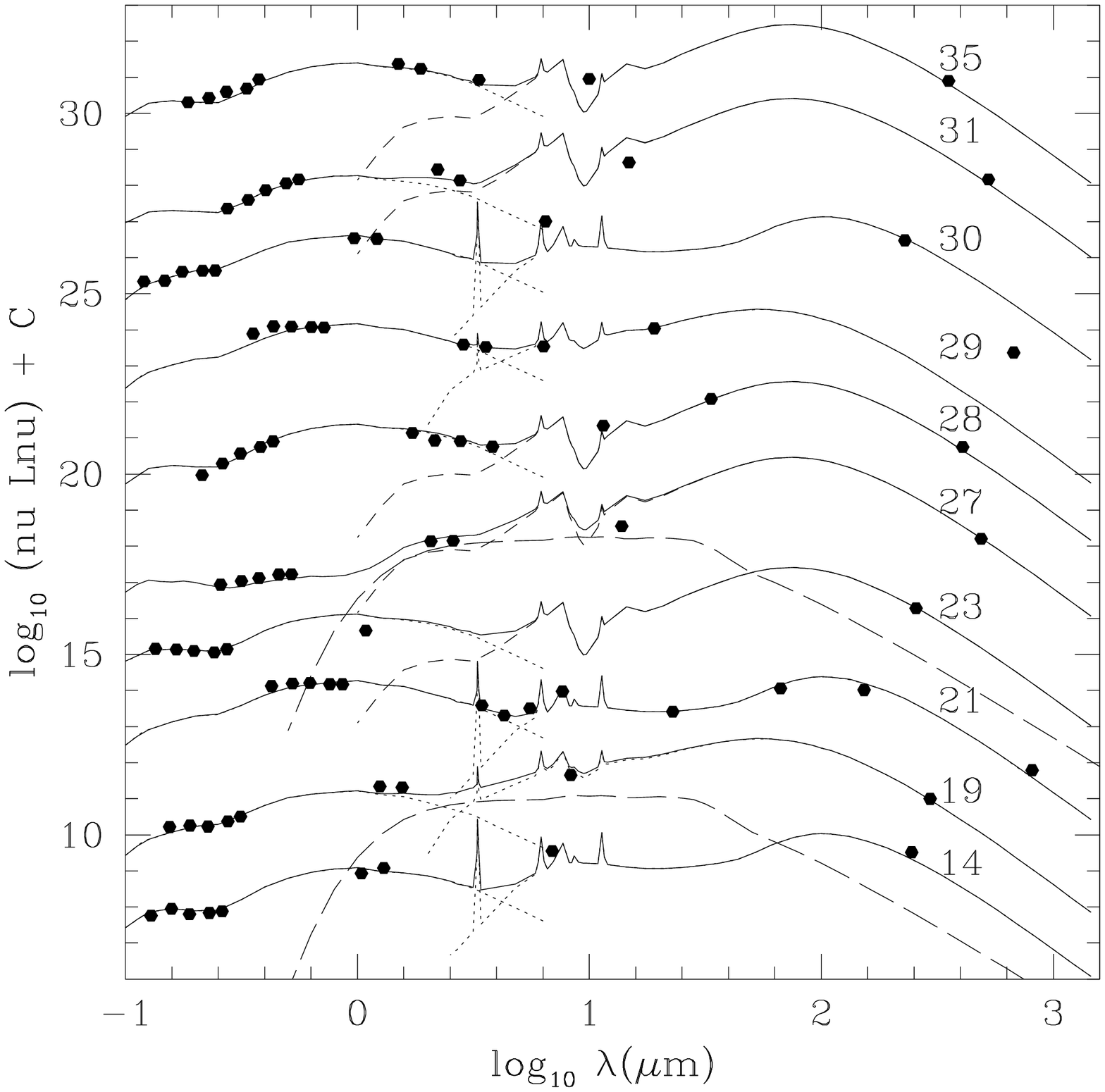}
\caption{LH: Seds of SWIRE sources with IRS data.
RH: Seds of SWIRE-SHADES sources in SXDS.
}
\end{figure}

\section{Photometric redshifts in SWIRE-VVDS, Lockman and ELAIS-N1}

Rowan-Robinson et al (2005) reported a photometric redshift analysis for two SWIRE samples, the 0.3 sq deg area of 
Lockman-VF and 6.5 sq deg in ELAIS-N1, based on the methods of Rowan-Robinson (2003), Babbedge et al (2004).
Here we show a comparison of $log_{10} (1+z_{phot})$ with
$log_{10} (1+z_{spect})$ for the SWIRE-VVDS sample (Fig 2(L)), which yields over 1300 reliable spectroscopic redshifts for 
SWIRE galaxies (Lefevre et al 2005, Ilbert et al 2006).
and for the whole SWIRE-Lockman area (Fig 2 (R)), in which we have carried out a substantial programme of spectroscopy
with Keck, Gemini-N and WIYN (Smith et al, 2006, in prep., Owen et al, 2006, in prep.).

When including 3.6 and 4.5 $\mu$m data in the solution we find accuracies of $\sim 5\%$ in (1+z), compared
with 10 $\%$ using optical data, and a reduced number of outliers (2$\%$).  The number of optical photometric bands is 
clearly also a factor.  For much of the Lockman area we have only gri data available.

\begin{figure}
\plottwo{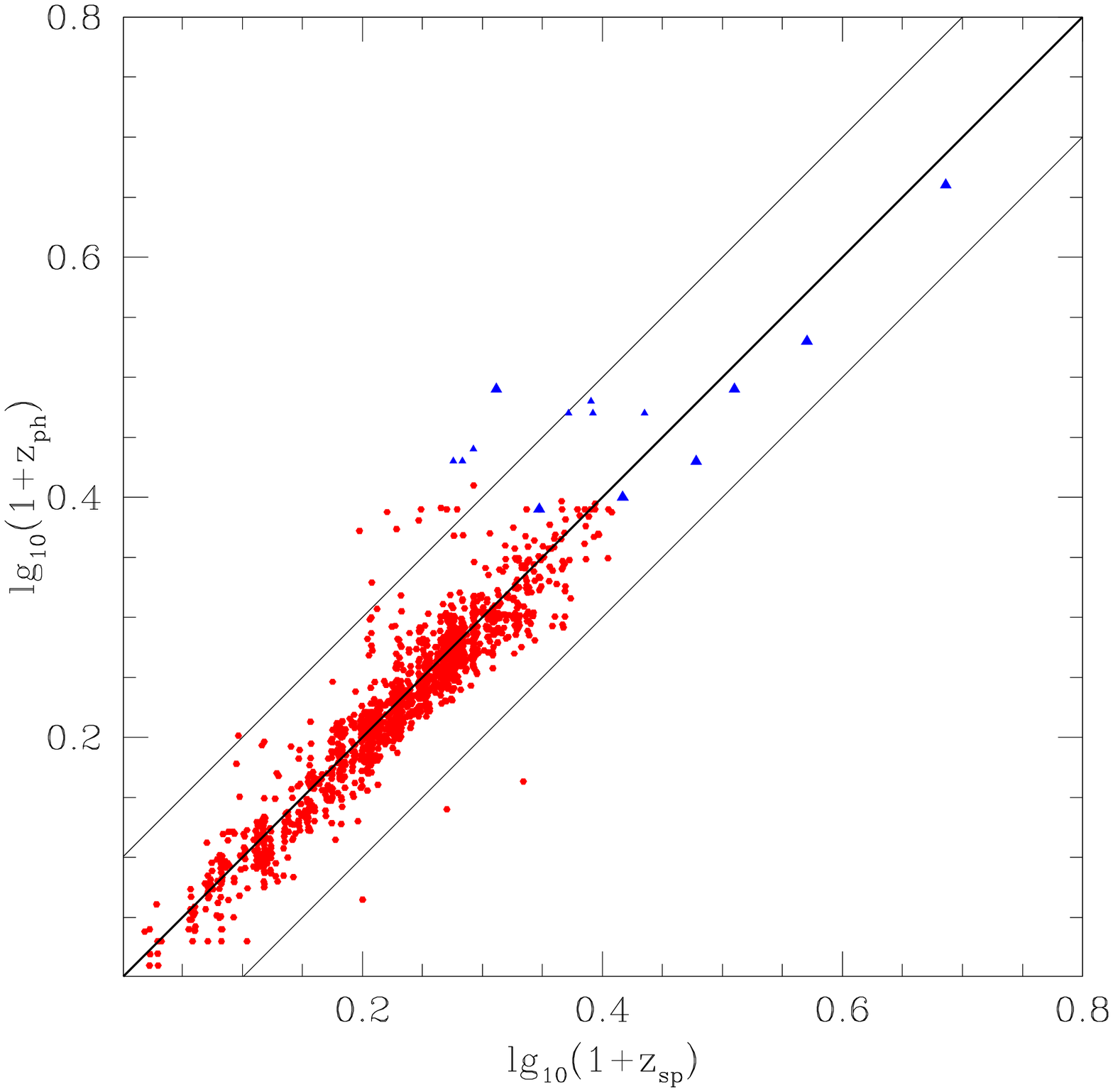}{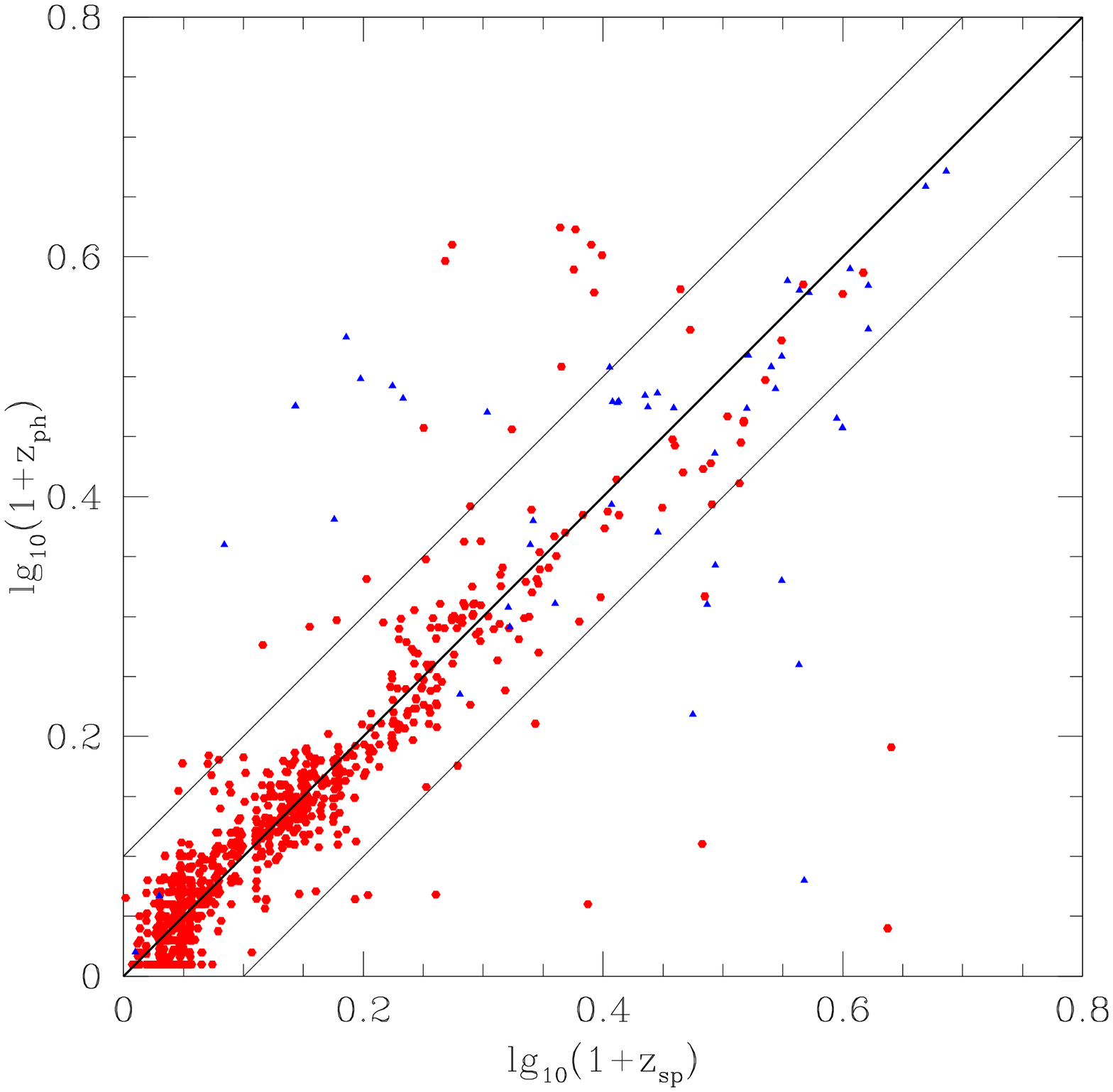}
\caption{LH: Photometric versus spectroscopic redshift for SWIRE-VVDS sources, showing good accuracy for galaxies out to z =
1.5.
RH: Photometric versus spectroscopic redshift for SWIRE-Lockman sources.
}
\end{figure}

\section{Redshift distributions}
Figure 3 (L) shows the redshift distributions for SWIRE-N1 sources with S(3.6) $>$
10 $\mu$Jy, above which the SWIRE survey is relatively complete,
with a breakdown into elliptical, spiral + starburst and quasar seds based on the photometric redshift fits.  
Subsequent to the analysis of Rowan-Robinson et al (2005), we have (1) improved that processing of AGN to
include the possibility of QSOs with significant extinction, (2) relaxed the requirements on the code to
take account of sources with very weak or no optical data (by using IRAC 5.8 and 8 $\mu$m data for these sources).
For comparison we show in the top panel the predictions
of Rowan-Robinson (2001).  Ellipticals cut off sharply at z $\sim$ 1.4.  Spirals also cut off at $\sim$ 1,5
but there is an extended tail of sources to z $\sim$ 4. 10$\%$ of SWIRE galaxies have z$>$2 and 4$\%$ have z$>$3.

The photometric estimates of redshift for AGN are more uncertain than those for galaxies, due to aliassing problems,
but the code is effective at identifying Type 1 AGN from the optical and near ir data.  
For some quasars there is significant torus dust emission in the 3.6 and 4.5 $\mu$m bands,
so inclusion of these bands in photometric redshift determination can make the fit worse rather than better.
We have therefore omitted the 3.6 and 4.5 $\mu$m bands if S(3.6)/S(r) $>$ 3.
Note that only 5 $\%$ of SWIRE sources are identified by the photometric redshift code as Type 1 AGN, and
of these only 5$\%$ are found to have $A_V > 0.5$.  

\begin{figure}
\plottwo{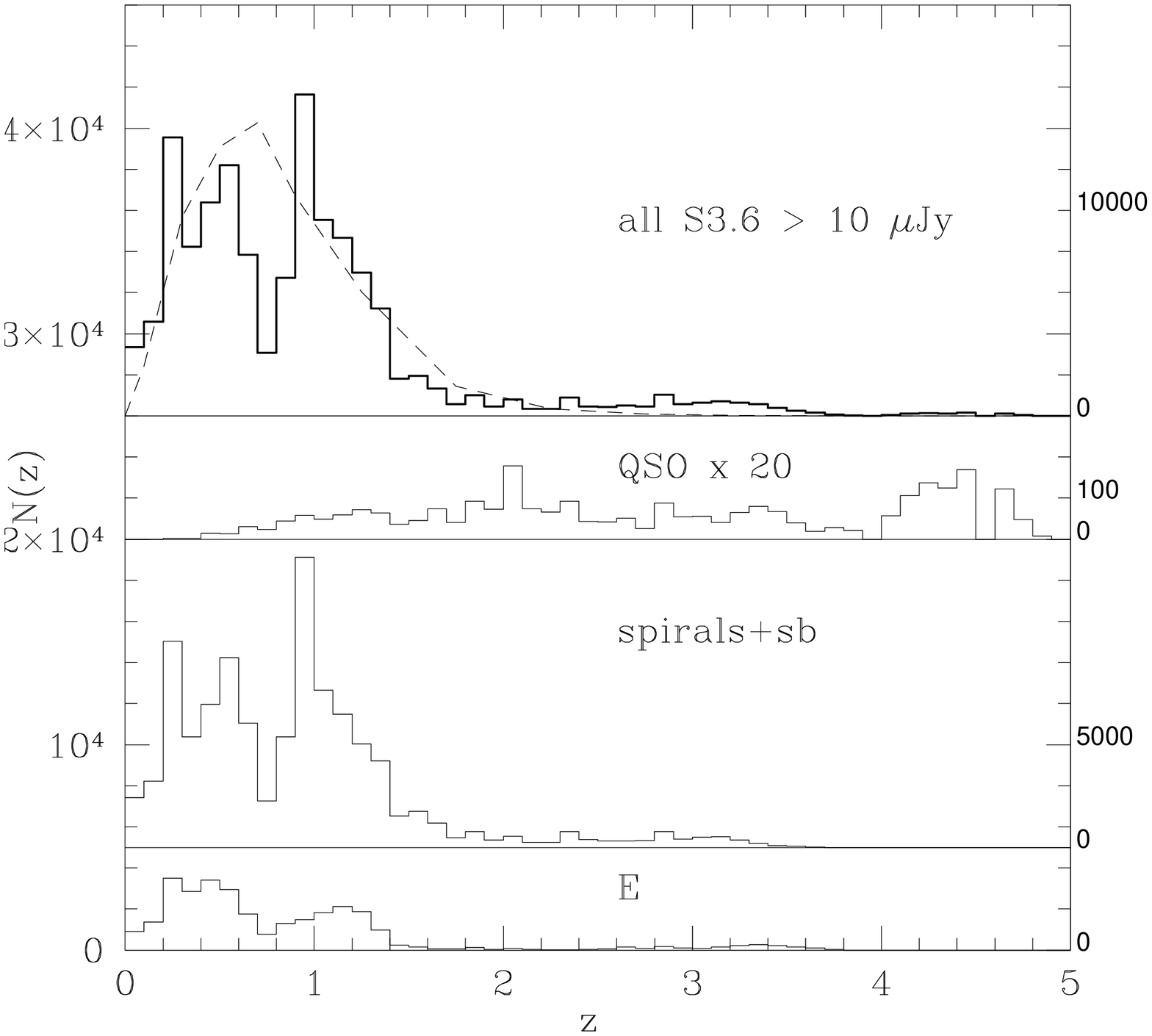}{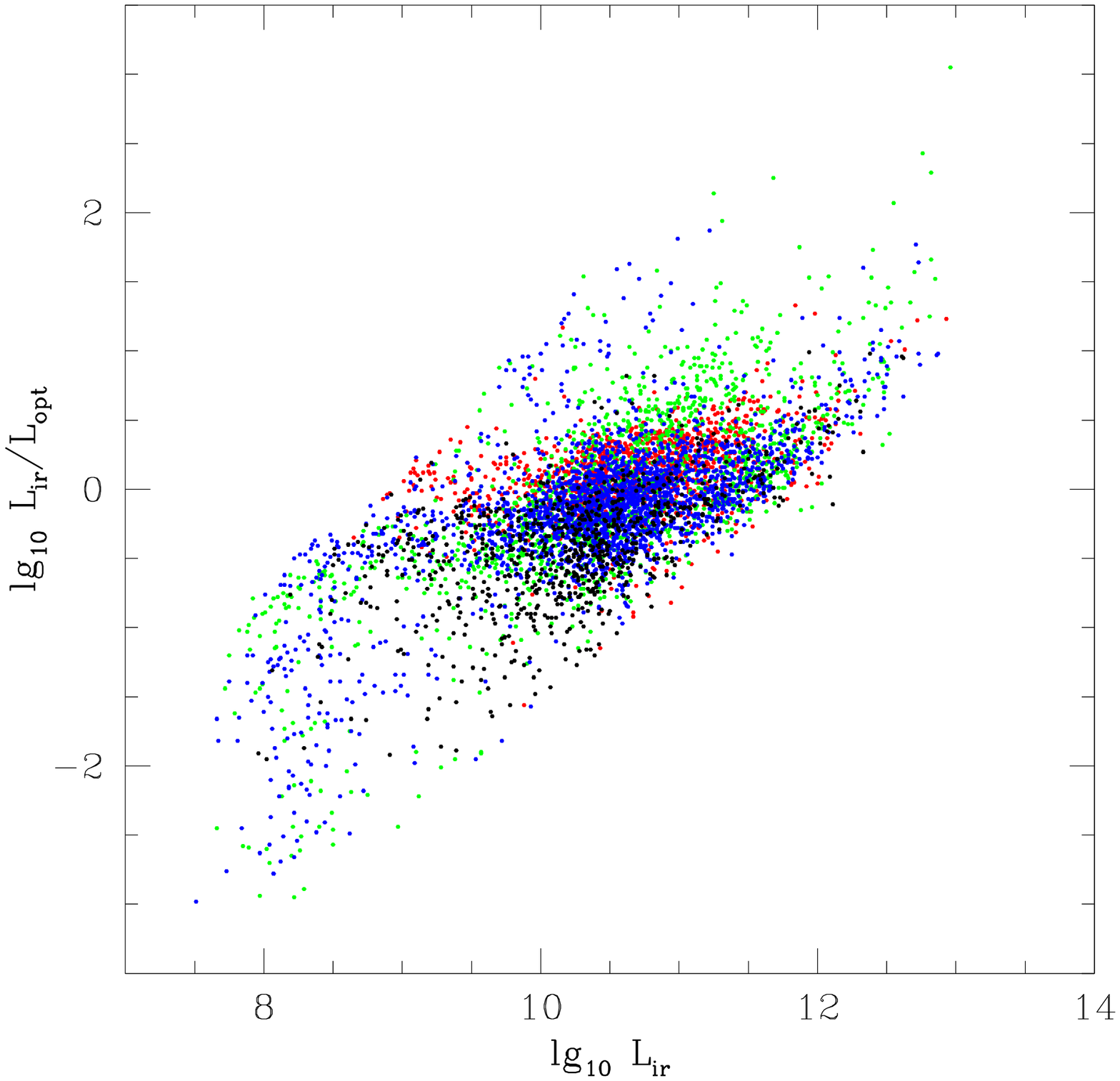}
\caption{LH: Photometric redshift histogram for SWIRE ELAIS-N1 sources with good 4-band optical IDs, and S(3.6) $>$ 10 $\mu$Jy.  
Top panel: all sources, broken curve: prediction of Rowan-Robinson (2001);
lower panels: separate breakdown of contributions of ellipticals, spirals
(Sab+Sbc+Scd)+ starbursts (Sdm+sb), and quasars.  The histogram for quasars has
been multiplied by 20.
RH: $L_{ir}/L_{opt}$ versus $L_{ir}$ for cirrus galaxies, colour-coded by optical template type
(black: E, red: Sab, blue: Sbc/Scd, green: Sdm,sb).
}
\end{figure}

\begin{figure}
\plottwo{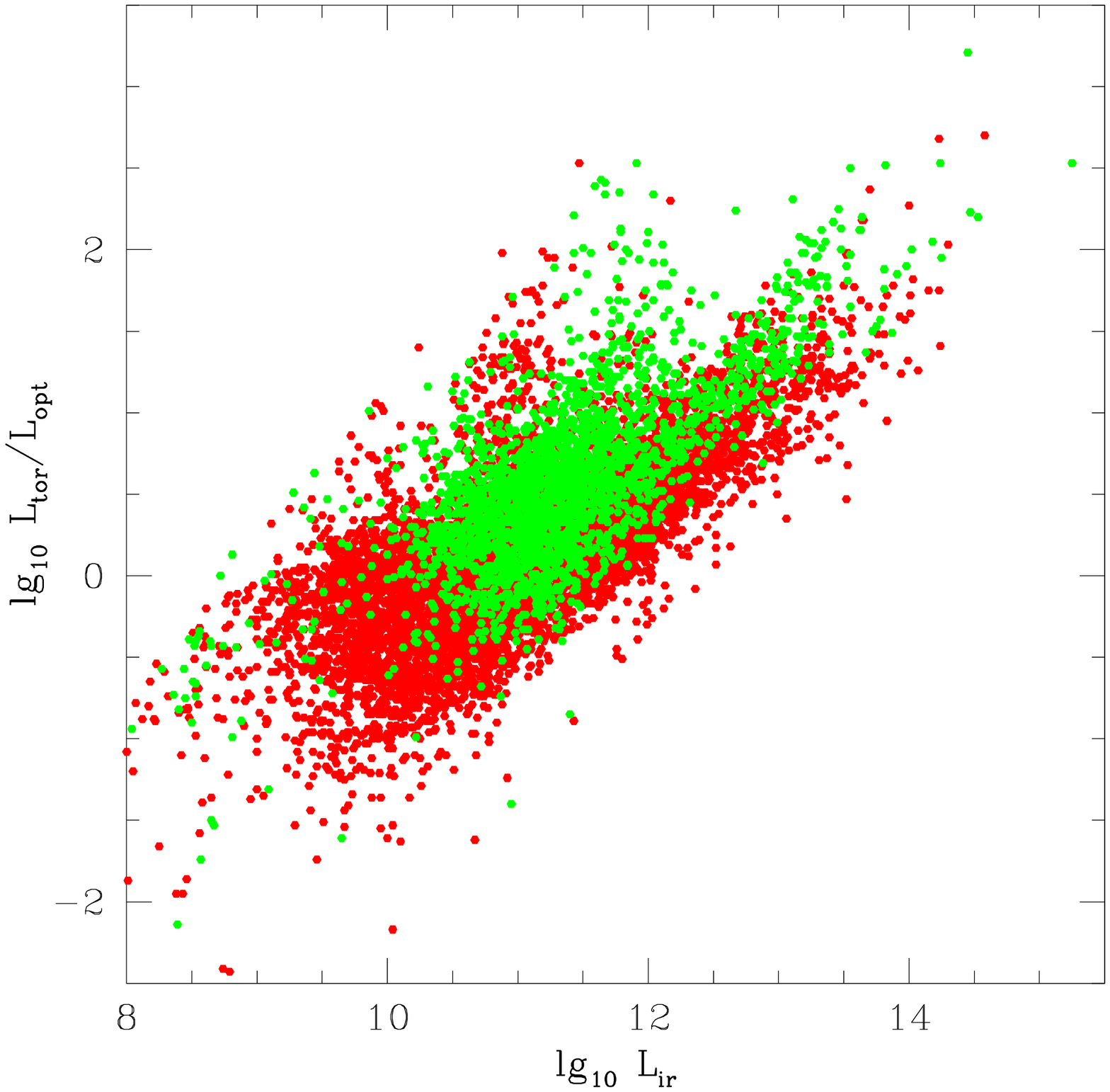}{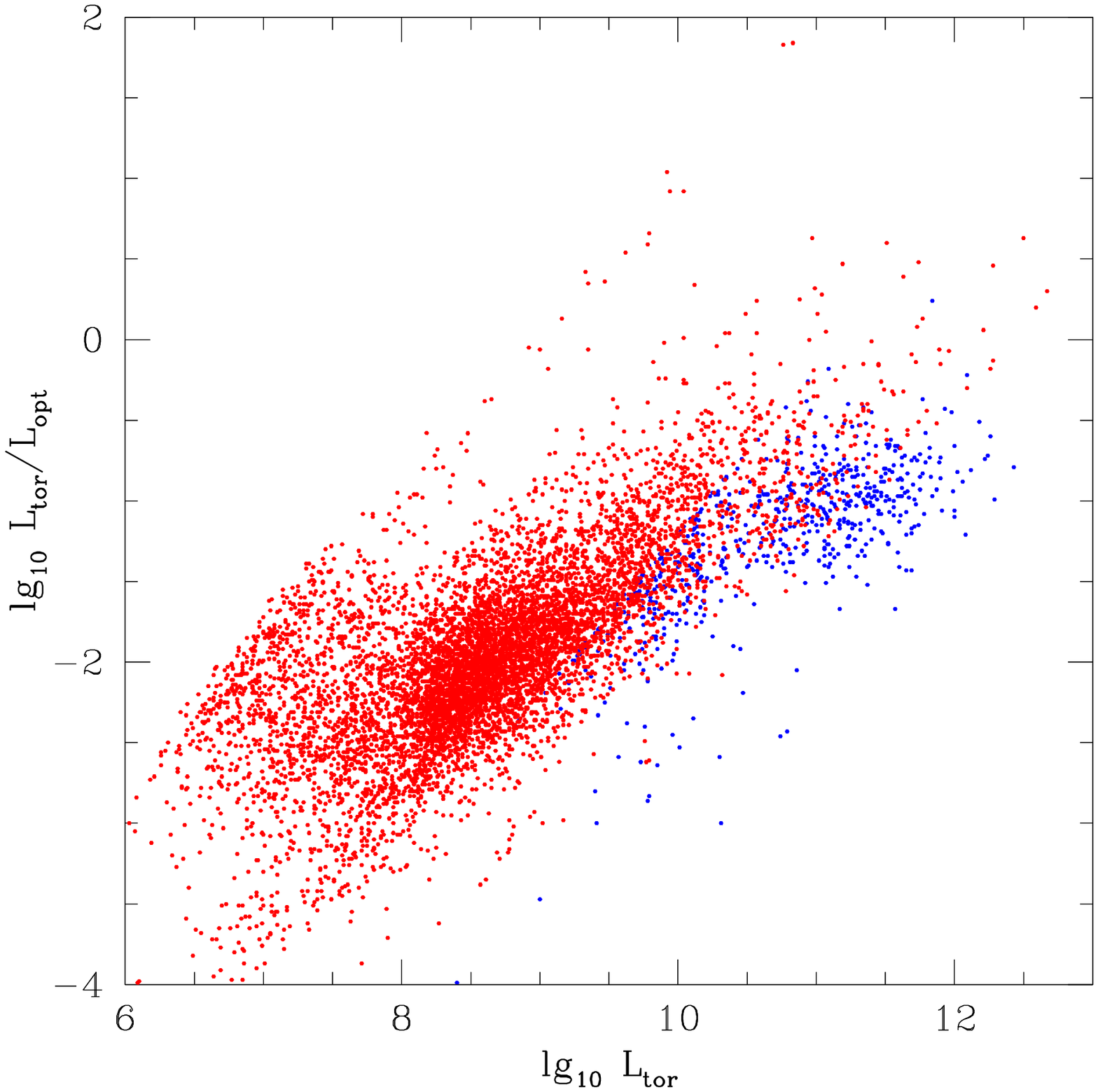}
\caption{LH: $L_{ir}/L_{opt}$ versus $L_{ir}$  for starburst galaxies (red: M82, green: A220).
RH: $L_{tor}/L_{opt}$ versus $L_{tor}$  for AGN dust tori (red: galaxy optical seds, blue: QSO optical seds).
}
\end{figure}

\section{Bolometric infrared and optical luminosities}
For sources detected at 70 or 160 $\mu$m, or with an infrared excess at 4.5-24 $\mu$m, relative to the template used for photometric redshift fitting, 
in at least two bands (one of which we require to be 8 or 24 $\mu$m),
we have determined the best-fitting out of cirrus, M82 starburst, Arp220 starburst or AGN dust torus 
infrared templates (cf Rowan-Robinson et al 2005). From the experience of modeling IRAS sources
(eg Rowan-Robinson and Crawford 1989), confirmed by the sed modeling described in section 3,
we allow (a) a mixture of cirrus and M82 templates, (b) a mixture of AGN dust torus and M82 templates, or
(c) an Arp 220 template (using a mixture of AGN dust torus plus A220 template did not improve the $\chi^2$ distribution).
 
We can estimate the bolometric luminosity corresponding to the infrared template 
and to the optical template used for photometric redshift determination.
Figure 3(R) and 4  shows the ratio of bolometric infrared to optical luminosity, $lg_{10}(L_{ir}/L_{opt})$, versus bolometric infrared luminosity,
 $lg_{10} L_{ir}$
for cirrus, starburst and AGN dust tori.

For cirrus galaxies with $L_{ir} < 10^{10} L_{\odot}$, a significant fraction of the infrared emission is
reemission of starlight absorbed by (optically thin) interstellar dust, so $L_{ir}/L_{opt}$ should
be interpreted as the optical depth of the interstellar dust.  Many very low values of $L_{ir}/L_{opt}$
($<$ 0.2) are due to elliptical galaxies with a small amount of star-formation.
However there is also an interesting population of luminous and ultraluminous cirrus galaxies, with $L_{ir} > L_{opt}$
(Rowan-Robinson et al 2004, 2005).
7 $\%$ of the cirrus galaxies identified in the N1 area have $L_{ir} > 3.10^{11} L_{\odot}$ and 2 $\%$ have
$L_{ir} > 10^{12} L_{\odot}$.  The implications are that (1) the quiescent phase of star formation was significantly
more luminous in the past (as assumed in the count models of Rowan-Robinson 2001), (2) the dust opacity of
the interstellar medium in galaxies was higher  at z $\sim$ 1, as expected from galaxy models with
star-formation histories that peak at z = 1-2 (Pei et al 1999, Calzetta and Heckaman 1999, Rowan-Robinson 2003).

For star-forming galaxies the parameter $L_{ir}/L_{opt}$ can be interpreted as approximately $10^{-9} \dot{M_*}/M_* (yr^{-1}$,
since $L_{ir} \sim 10^{10} \dot{M_*}$ and $L_{opt} \sim 10 M_*$,
i.e. as $10^{-9} \tau^{-1}$, where $\tau$ is the time-scale in yrs to accumulate the present stellar mass, forming
stars at the current rate ($\dot{M_*}/M_* = \tau^{-1}$ is the specific star formation rate). 
The galaxies with Arp 220 templates tend to have high values of $L_{ir}/L_{opt}$, consistent
with the idea that they have lower $\tau$.  M82 type starbursts are less extreme and range over rather similar 
values of $L_{ir}$ and  $L_{ir}/L_{opt}$  to the cirrus galaxies.

For AGN dust tori (blue points in Fig 5R), the ratio $L_{ir}/L_{opt}$ can be interpreted (for Type 1 AGN) as the 
covering factor of the torus.  Most have values in the range 0.03-0.5, with a median value $\sim$ 0.1.
If we restrict attention to sources with $L_{ir} > 10^{11} L_{\odot}$, we find that one third have QSO optical seds,
while two thirds have galaxy optical seds, implying that the Type 1: Type 2 ratio is 2:1.  
Further discussion of SWIRE quasars is given by Franceschini et al (2005), Hatziminaoglou et al (2005, 2006), 
Polletta et al (2006).

5$\%$ of 24 $\mu$m sources have extremely high infrared luminosities, in the hyperluminous class
($> 10^{13} L_{\odot}$).  Clearly spectroscopy is needed to determine these redshifts more accurately.

The ir templates can be used to predicted fluxes at longer wavelengths.  Fig 5 (L) shows
predicted fluxes at 70 $\mu$m, derived from template fits to 3.6-24 $\mu$m data, compared
with the observed MIPS fluxes.  The agreement is remarkably good.  Fig 5 (R) shows predicted fluxes
at 350 $\mu$m versus redshift for SWIRE-N1 sources.

\section{Source-counts and luminosity functions.}

Figures 6 and 7 show the differential counts at 24 and 160 $\mu$m observed by SWIRE
(Shupe et al 2006, Afonso Luis et al 2006), compared with new models developed from
the approach of Rowan-Robinson (2001). 
These new models allow separate rates of evolution for each component;  while similar histories
are found for cirrus, M82 starbursts and AGN dust tori, the Arp 220 population requires significantly
steeper evolution to fit the counts, implying a more dramtic evolution for these extreme objects.

\begin{figure}
\plottwo{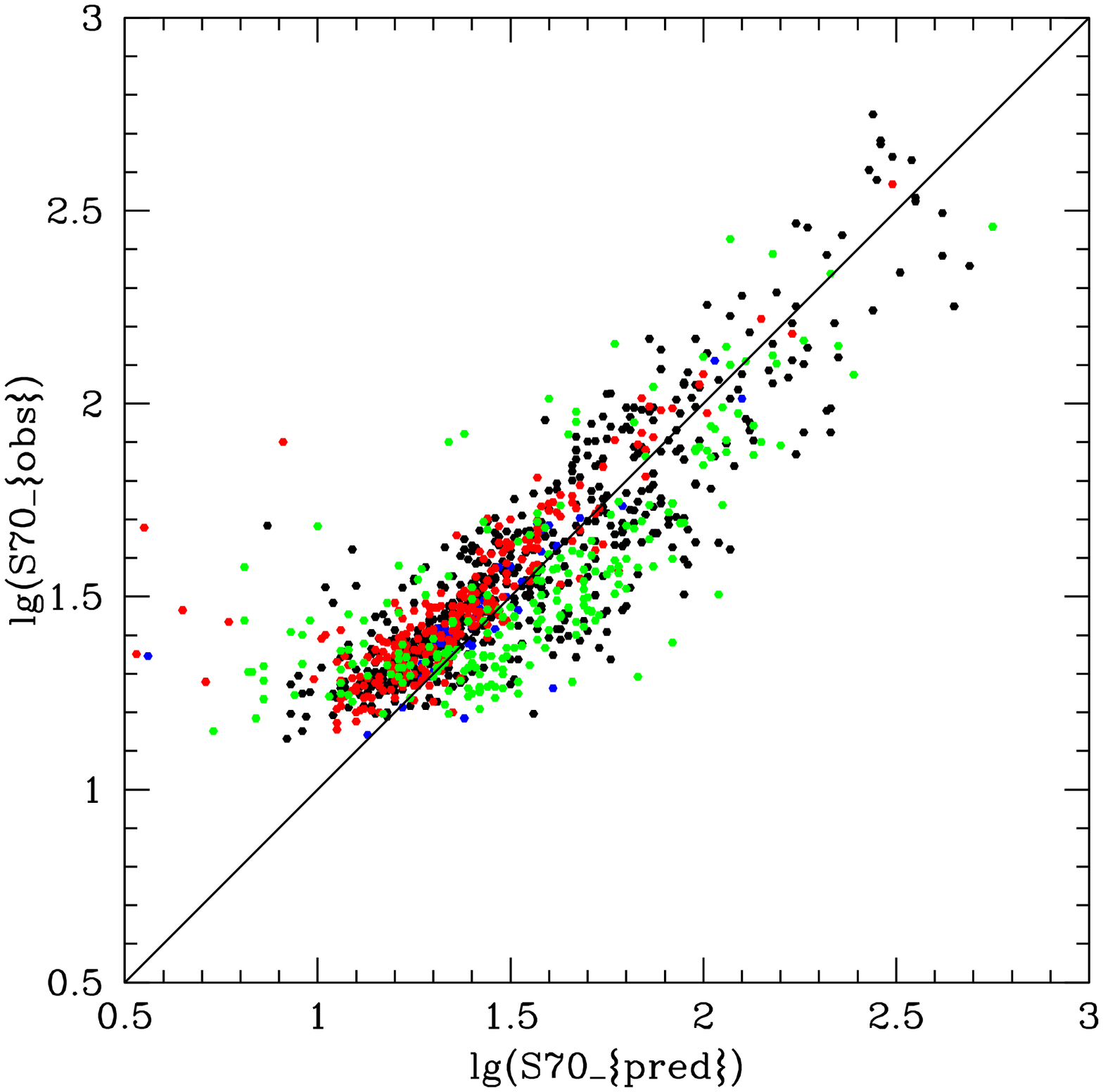}{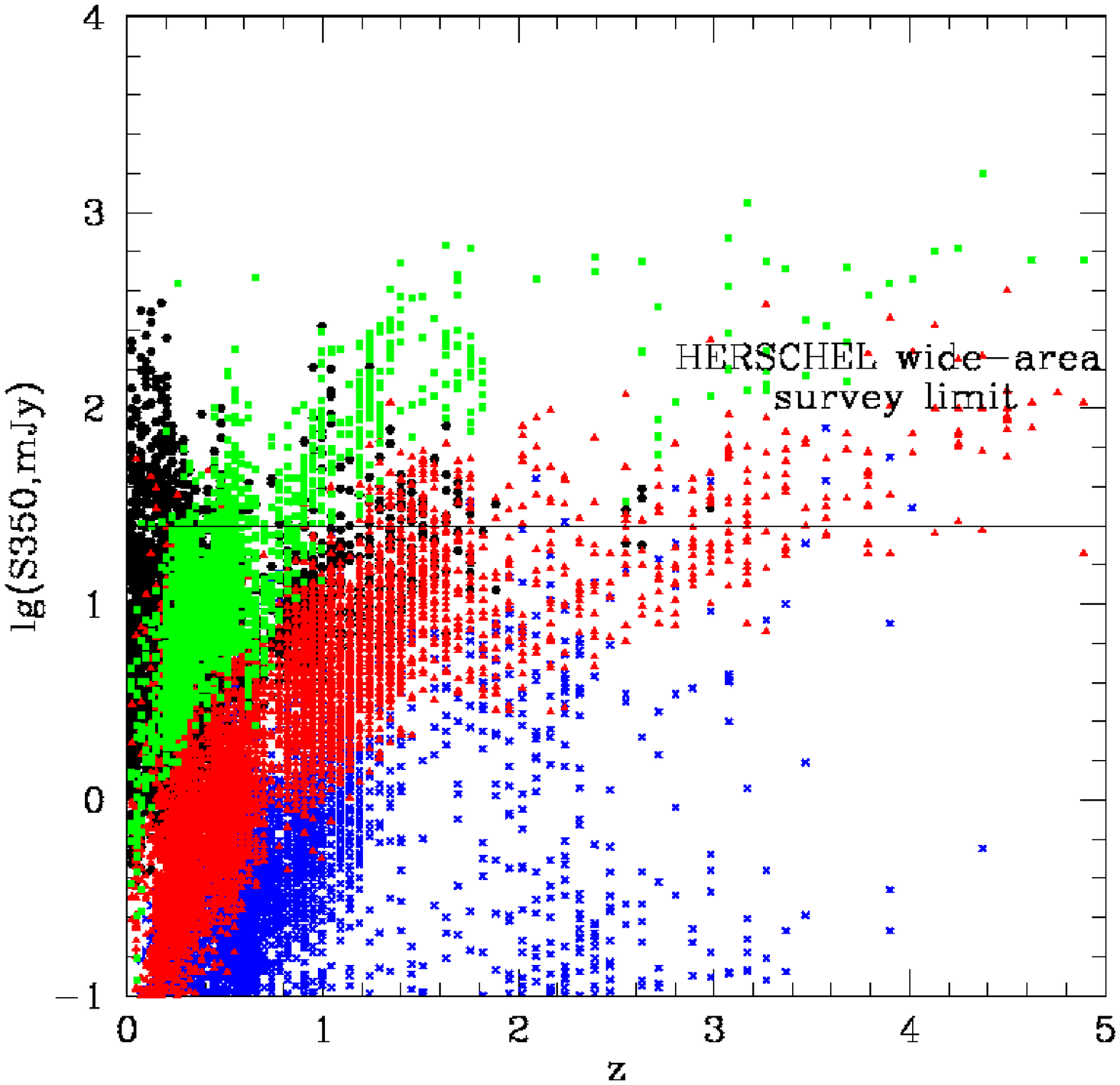}
\caption{LH: Predicted versus observed 70 mu flux.
RH: Predicted 350 $\mu$m flux versus redshift for SWIRE-N1 sources.  The
limit of a planned wide-area survey with HERSCHEL-SWIRE is indicated.  Black: cirrus, red: M82 starburst, green: 
A220 starburst, blue: AGN dust torus.
}
\end{figure}

\begin{figure}
\plottwo{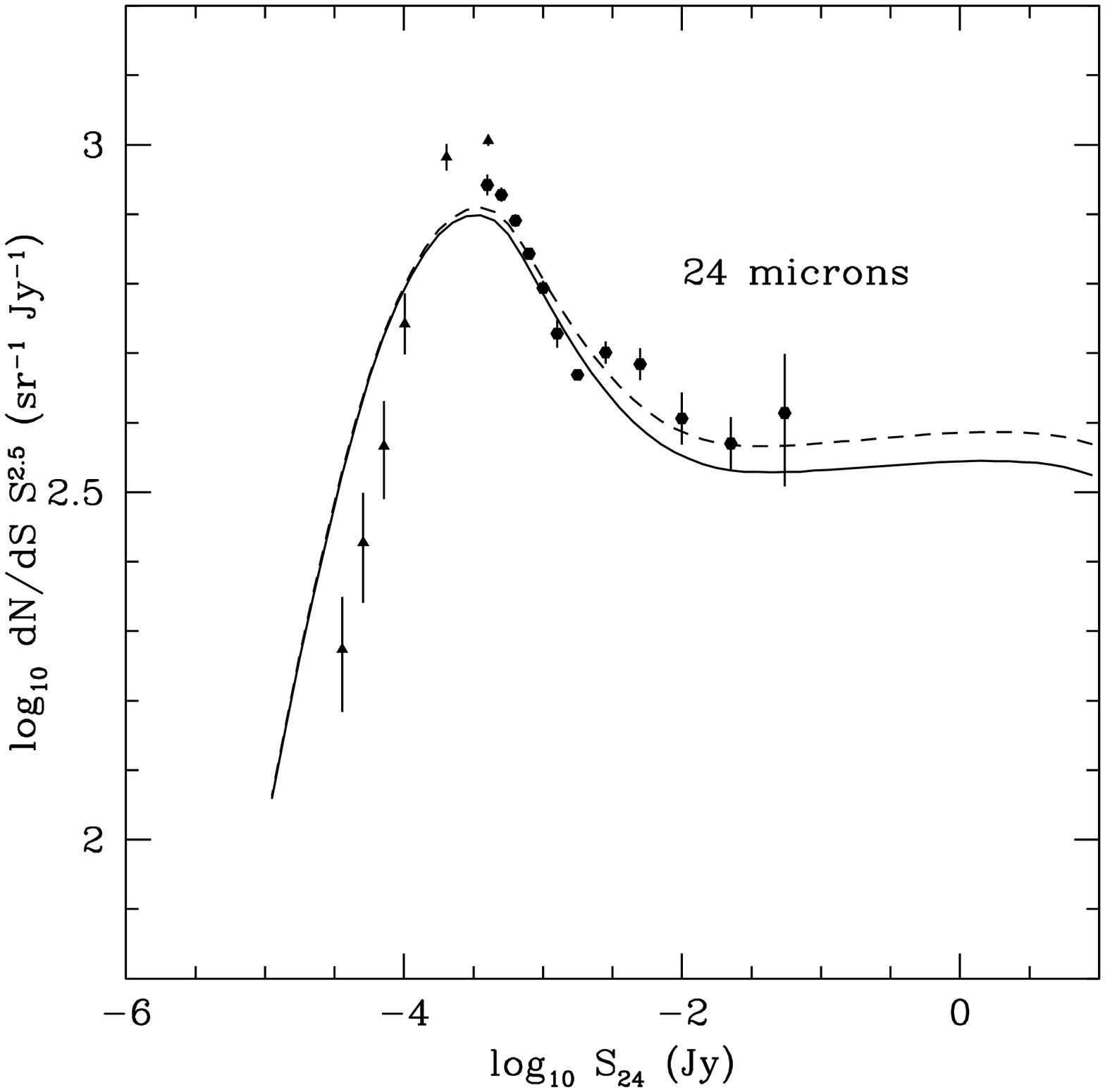}{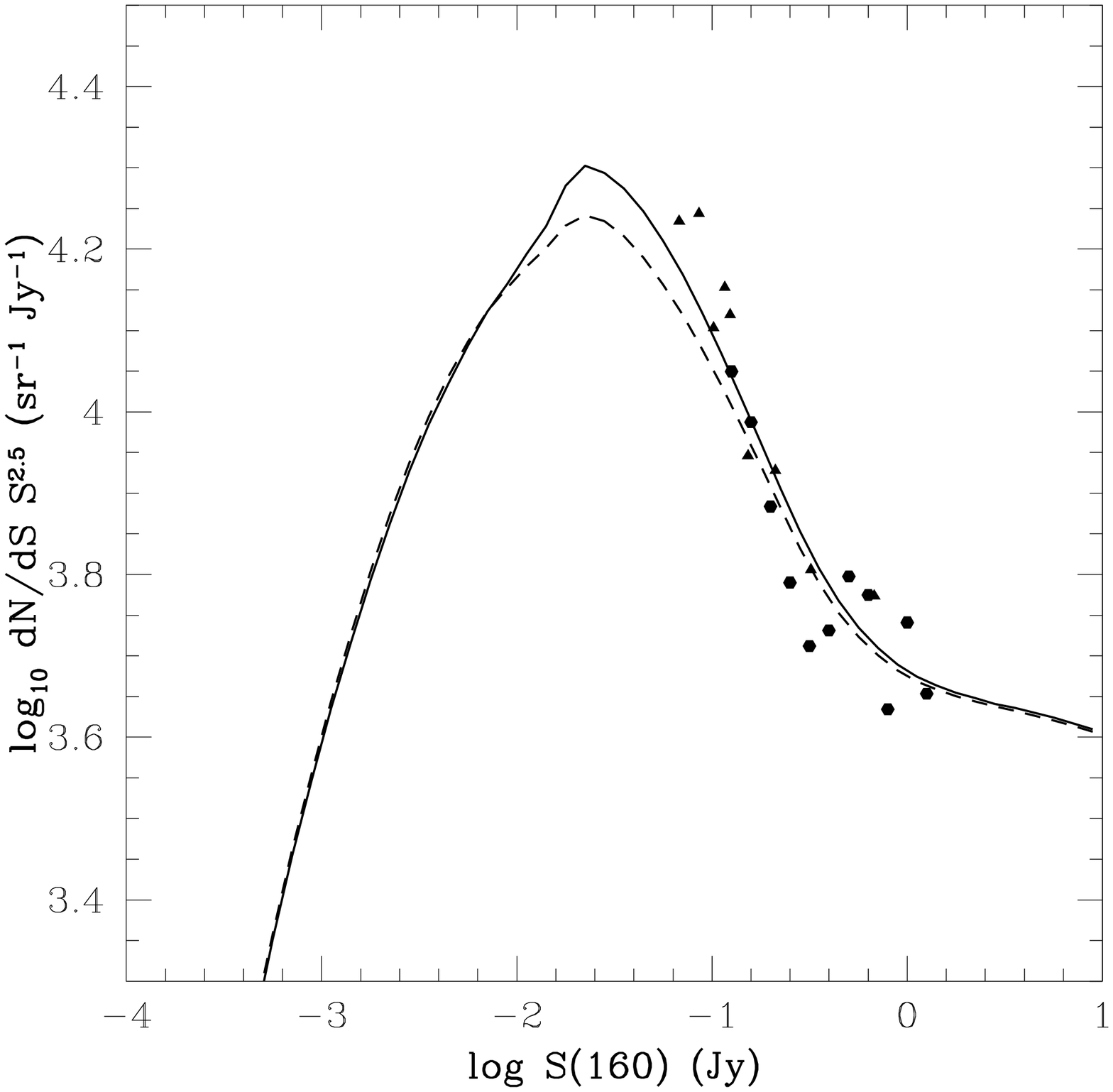}
\caption{LH: 24 $\mu$m differential counts($mJy^{1.5} deg^{-2}$).  RH: 160 $\mu$m differential counts ($Jy^{1.5} sr^{-1}$).
Filled circles are SWIRE data, filled triangles are from Papovich et al (2004) (LH) and
Frayer et al (2006) (RH).  The curves are new models based on the approach of Rowan-Robinson (2001). 
}
\end{figure}

Babbedge et al (2006) have derived luminosity functions at 3.6-24 $\mu$m, using the SWIRE-N1
photometric redshifts (Fig 8).  The evolution of the 3.6 $\mu$m luminosity function is consistent
with the passive evolution of starlight, but the evolution at 24 $\mu$m implies strong
evolution of the star-formation history between z = 0 and 1.

\begin{figure}
\plottwo{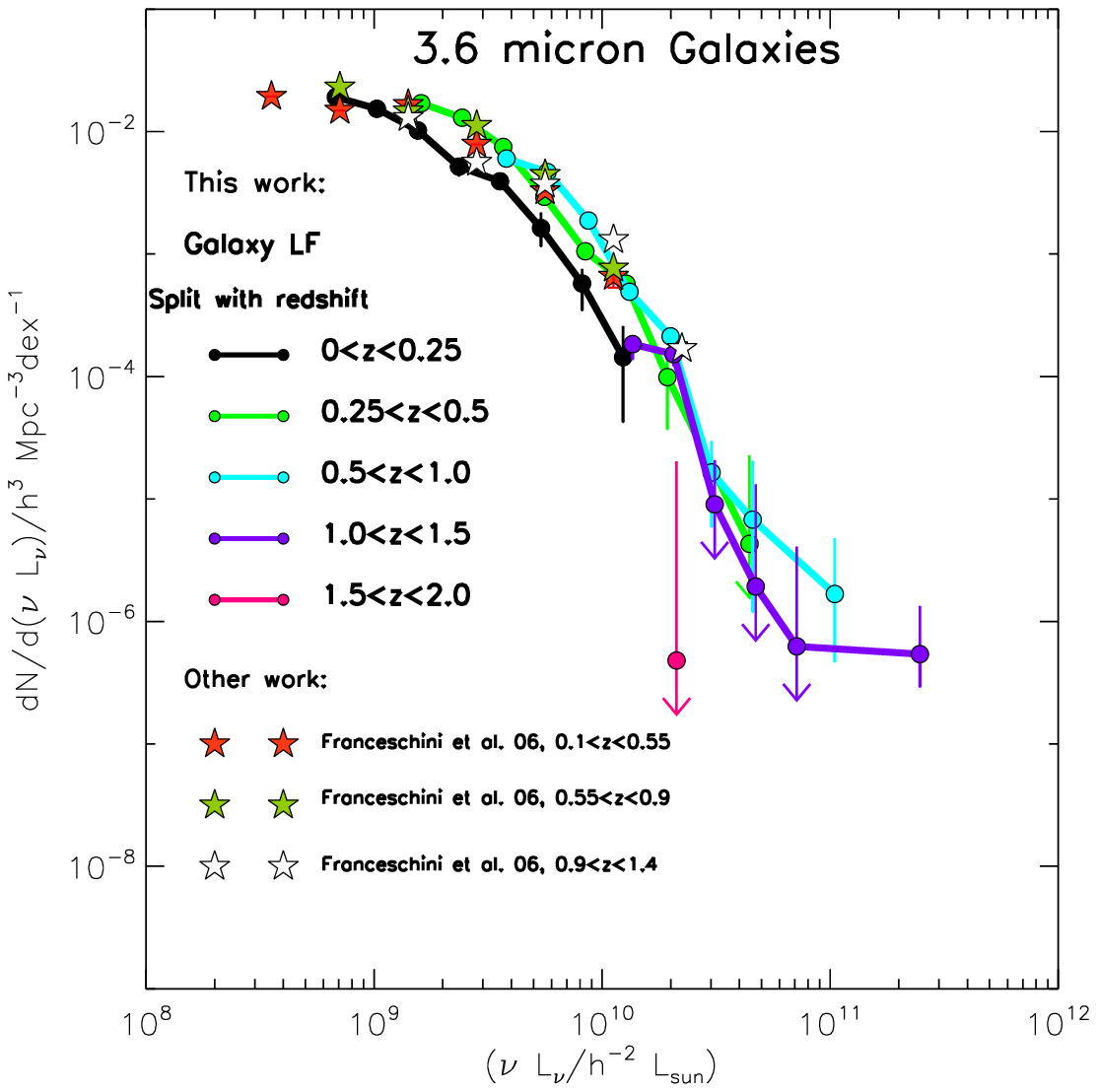}{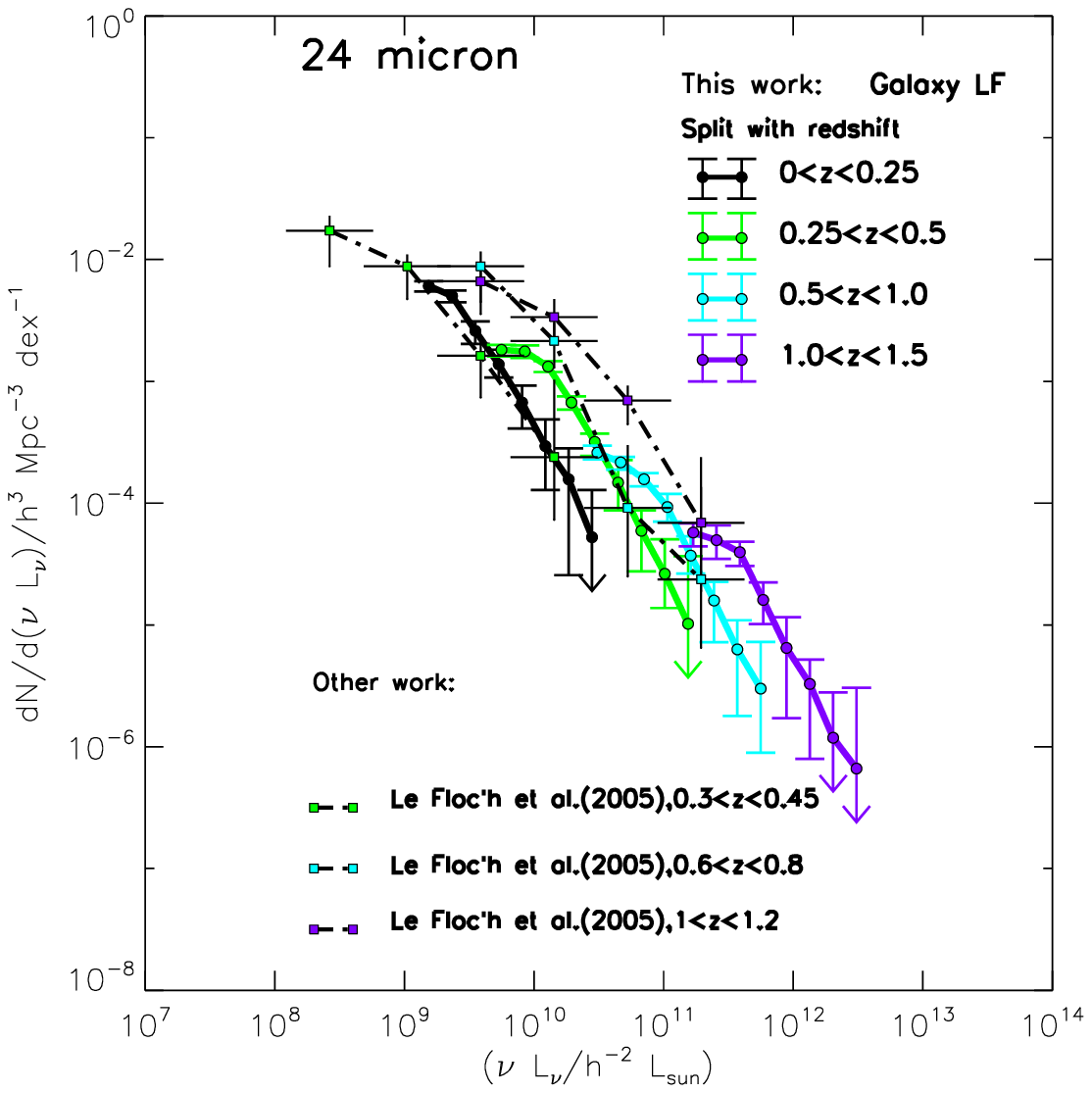}
\caption{LH: Luminosity function at 3.6 $\mu$m for different redshift bins.
RH: Luminosity function at 24 $\mu$m.
}
\end{figure}



\end{document}